\documentclass[usenatbib,usegraphicx]{mn2e}
\usepackage{longtable}
\usepackage[english]{babel}

\title[Search for LBV Candidates in the M\,33 Galaxy]{Search for LBV Candidates in the M\,33 Galaxy}
\author[Valeev, Sholukhova, Fabrika]{A.F. Valeev, O. Sholukhova, and S. Fabrika
\\
Special Astrophysical Observatory, Nizhnij Arkhyz 369167, Russia\\
E-mail: azamat@sao.ru, olga@sao.ru, fabrika@sao.ru}

\begin{document}
\maketitle

\newcommand{\Ha}{H$\alpha$}
\newcommand{\SII}{[SII]}
\newcommand{\OIII}{[OIII]}
\newcommand{\BVlim}{0\fm{}35}
\newcommand{\Vlim}{18\fm{}5}

\newcommand{\aaps}{Astronom. and Astrophys. Suppl.}
\newcommand{\aap}{Astronom. and Astrophys.}
\newcommand{\aj}{Astronom. J.}
\newcommand{\apj}{Astrophys. J.}
\newcommand{\apjs}{Astrophys. J. Suppl.}
\newcommand{\apss}{Astrophys. and Space Sci.}
\newcommand{\mnras}{Monthly Notices Roy. Astronom. Soc.}
\newcommand{\pasp}{Publ. Astronom. Soc. Pacific}
\newcommand{\apjl}{\apj}
\selectlanguage{english}

\begin{abstract}
A total of 185 luminous blue variable star (LBV) candidates with
V $<$ \Vlim{} are selected based on the results of aperture
photometry. The primary selection criterion is that the
prospective candidate should be a blue star with  \Ha{} emission.
In order not to overlook appreciably reddened LBV candidates, we
compose an additional list of 25 red (\BVlim{} $<$ $B-V$ $<$ 1\fm{}2,
$V$ $<$ \Vlim{}) emission star candidates. A comparison with the list
of known variables in the  M\,33 galaxy showed 29\% of our
selected candidates to be photometrically variable. We also find
our list to agree well with the lists of emission-line objects
obtained in earlier papers using different methods.
\end{abstract}

\section{INTRODUCTION}
Luminous blue variables are the most massive stars during the
final stage of evolution~[[\cite{HD1994_LBVreview}]. It is a
rather short-lived stage characterized by high mass-loss rate and
mass ejections into the interstellar medium during outbursts. The
bolometric luminosities of such stars are close to the Eddington
limit, below which radiation pressure can still be balanced by
gravity forces.

Modern model computations often involve explicitly preset time of
the beginning of the LBV evolution and the duration of this stage
(see, e.g.,~[\cite{Meynet2007}]), because these parameters
are impossible to infer via self-consistent modeling. The input
parameters in these cases are based on the most recent
observational data. Modeling is further complicated due to the
lack of a consensus concerning the evolutionary sequence of
massive stars -- e.g., even the LBV $\to$ WR transition remains an
open issue
([\cite{Smith_Owocki06}], [\cite{ContiWNHstars}], [\cite{Smith08}]).
The number of known LBV stars in our Galaxy is too small to test
the agreement between the model tracks and observational data.

The number of known and well-studied massive stars at the final
stages of evolution --- LBV stars, WR stars, B[e]-type
supergiants, and supergiants and hypergiants with various
temperatures --- should be increased considerably. These objects
have very different observational manifestations. Only with a
sufficient number of stars with known parameters it will be
possible to reliably identify the results of modeling of the
evolution of massive stars with observed objects. On the other
hand, a considerably increased sample will make the interpretation
of model computations more reliable. The discovery and study of
new massive stars at the final stages of evolution  (hereafter
referred to as ``LBV-like'' stars) would lay down the necessary
basis for linking theory with observation so that the studies of
LBV stars will no longer remain mostly descriptive.

The main goal of this paper is to search for and identify LBV candidates. In the Milky Way such
objects are hidden by strong interstellar extinction in the Galactic disk. They are currently
discovered in IR sky surveys (e.g.,~[\cite{IRsearchMS}]).Its fortunate orientation
and sufficiently large population of early-type stars makes the M\,33 galaxy of the Local Group  an
ideal object to be searched for  LBV-like stars~[\cite{M33Ivanov}]. Here we adopt an M33 distance
modulus of  24\fm9 (e.g., [\cite{M33Distance}]), which corresponds to a
distance of 950~kpc.

In their review, Humphreys and Davidson
[\cite{HD1994_LBVreview}] summarize all the data known about
LBV stars by that time: their list of confirmed LBV stars in M\,33
included Var\,B, Var\,C, Var\,2, and Var\,83, and  V\,532
(GR\,290, the ``Romano star''
 [\cite{Romano78}]) was considered to be an LBV candidate. Later, the latter had its
LBV status confirmed both spectroscopically
[\cite{Fabrika2000}] and
photometrically~[\cite{Kurtev_etal2001}]. It was studied in
more detail by Fabrika et al.~[\cite{Fabrika_etal2005}] and
Viotti et al.~[\cite{Viotti2006, Viotti2007}. The
star Var\,A in M\,33 is now commonly classified as a cool
hypergiant~[\cite{Humph_etal1987}], [\cite{Humphreys_etal2006}], [\cite{Viotti2006}].
Although this object exhibits all the features characteristic for
LBV stars --- it suddenly brightened in
1950--1953~[\cite{HubbleSandage53}] to become one of the most
luminous stars in  M\,33 with a spectral type F --- its current
spectral type (M) is untypical for classical LBV stars. However,
given the fewness of LBV stars hitherto studied, it is possible
that the known classical LBV states [\cite{HD1994_LBVreview}]
do not cover all possible properties of these objects. The Var\,A
star meets the criteria of LBV class both in terms of luminosity
(mass) and photometric variability~[\cite{NewLBVinM33}].

Spectroscopic observations of LBV-like star candidates from our
list allowed us to discover a new (the seventh)  LBV star N93351
in the M\,33 galaxy [\cite{NewLBVinM33}]. We use very limited
archival data to construct the light curve of the star and find it
to be variable with the light variations of about 0\fm4 a year.
Further observations of  N93351 both photometric and spectroscopic
are needed to confirm the evolutionary status of the star.

Various groups of authors~[\cite{
Neeseetal91}], [\cite{Spiller1992}], [\cite{Calzetti1995}], [\cite{FabShol95}], [\cite{MasseyUIT}], [\cite{
Shol_etal97}], [\cite{Shol_Fab2000}], [\cite{Corral_Herrero03}]
used various methods to search for LBV stars in M\,33. The
principal method consists in searching for \Ha{} sources
coincident with early-type stars. Although LBV stars need not
inevitably be blue objects~[\cite{Sterken_etal2008}], many
authors targeted their search on early-type stars. Some authors
looked for SS\,443-like
objects~[\cite{Neeseetal91}], [\cite{Calzetti1995}], [\cite{FabShol95}].
The spectrum of SS\,433 resembles those of late WR
stars~[\cite{Fab04}].

The team of Massey et.\,al [\cite{Massey2006}] made a new
step toward the study of the massive stellar population of the
M\,33 galaxy. The above authors used CCD images of the galaxy to
produce a catalog of 146622 stars down to a limiting magnitude of
23 containing broad-band photometric measurements with an accuracy
of 
{1--2\%}. In their next paper~[\cite{Massey2007_Ha}],
the team reported a list of emission stars in Local Group galaxies
including M\,33. The main purpose of this study was to find new
LBV star candidates. To this end, the above authors adopted the
following criteria:
a constraint on the \Ha{} line flux; \Ha{} line flux had to be greater than the [SII]
line flux; the [OIII] line flux had to exceed the continuum flux, and a constraint was imposed on the
star's color (see [\cite{Massey2007_Ha}] for details).
Massey et al.~[\cite{Massey2007_Ha}] report 37 LBV stars and LBV candidates in
M\,33. Some of the candidates reported by Massey et al.~[\cite{Massey2007_Ha}]
have been identified before [\cite{Corall1996}], [\cite{Shol_etal97}], [\cite{FabShol95}], [\cite{Shol_Fab2000}], [\cite{Fabrika2000}].

We performed independent aperture photometry of bright objects
with $V <$\Vlim{} in all filters (UBVRI and \Ha) on the CCD frames
of Massey et al.~[\cite{Massey2006}].
We used the same methods to select LBV candidates as Fabrika et
al.~[\cite{Fabrika_etal1997}]. Emission-line objects are
easily identifiable on the ``\Ha{}-line flux vs. V-band flux''
diagram, because they lie above the linear relation outlined by
objects without the \Ha{} emission  (see below for details).

As a result of this work we composed a list of blue LBV-like star candidates in  M\,33.
In order not to overlook highly reddened objects we composed an additional list of red
\Ha{} emission objects selected with less stringent color criteria.

\section{OBSERVATIONAL DATA}

We downloaded from the NOAO science archive \linebreak ({\tt
http://www.archive.noao.edu/nsa/}) all primarily reduced  M\,33
frames taken by Massey et al.~[\cite{Massey2006}]  with the
coordinate grid applied. Observations were made in October 2000
and September 2001 with the the 4-m telescopes of the KPNO and
CTIO telescopes using UBVRI filters and narrow-band (50~\AA)
\Ha{}. The UBVRI and \Ha{}-filter images were taken during
photometric nights when the seeing was 0\farcs{}6 to 0\farcs{}8,
and 0\farcs{}8 to 1\farcs{}0, respectively. The detector emplolyed
was a mosaic of eight 2048$\times$4096 CCDs. Each image has a
field of view of 36\arcmin{}$\times$36\arcmin and a scale of
0\farcs{}26/pix (after primary reduction and distortion
correction). The M\,33 galaxy was subdivided into three zones with
each zone observed five times with a small offset to fill the gaps
between the CCDs in the mosaic. In this paper we use 15 mosaic
images in each filter. A more detailed description of the data and
primary reduction steps can be found in [\cite{Massey2006}].

\section{APERTURE PHOTOMETRY}
\label{Sect03}

Our aim was to select stars with \Ha{} emission --- the LBV
candidates. The \Ha{} images contain both point and extended
objects and therefore in the process of photometry we must check
that measurements in different filters refer to the same object.
This condition should be satisfied even in crowded fields and in
complex \Ha{} regions, something that is difficult to achieve in
automatic photometry of several hundred thousand objects
[\cite{Massey2007_Ha}].

Aperture photometry has an  important advantage over PSF
photometry in that it makes it possible to determine the flux from
an object without making any assumptions about its structure and
the form and parameters of the point-spread function (PSF) in the
Earth atmosphere. The use of the averaged PSF may introduce an
error in the estimated flux due to the variation of the form of
the PSF across the field. A problem with aperture photometry in
crowded fields is that the aperture radius should be chosen
individually for each object so as to keep all other objects
outside the aperture. The determination and application of
aperture correction is yet another problem to be faced when
performing aperture photometry. However, if we need not  determine
the total flux, but only the flux difference in different filters,
the same aperture size can be used in all filters. It is also
important that seeing is approximately the same in all filters.
The observations of Massey et al.~[\cite{Massey2006}] meet
this condition.

We used standard IRAF\footnote{{IRAF} is distributed by the
National Optical Astronomy Observatory, which is operated by the
Association of Universities for Research in Astronomy (AURA) under
cooperative agreement with the National Science Foundation.} tools
for photometric measurements and all performed auxiliary
operations in the batch mode using programs written in Python
language.

\begin{figure}%[]
\includegraphics{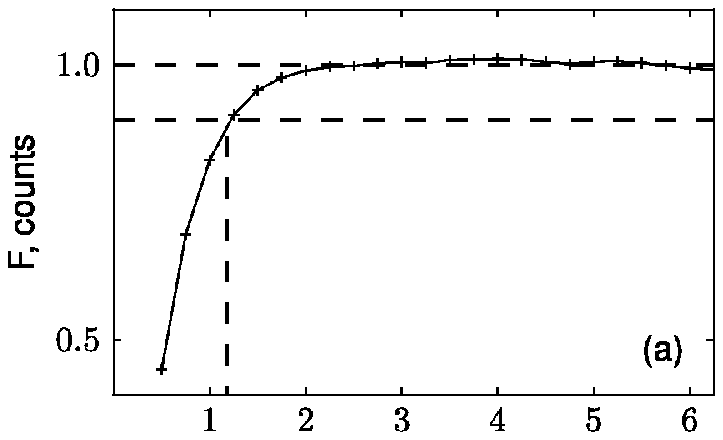}\vfil
\includegraphics{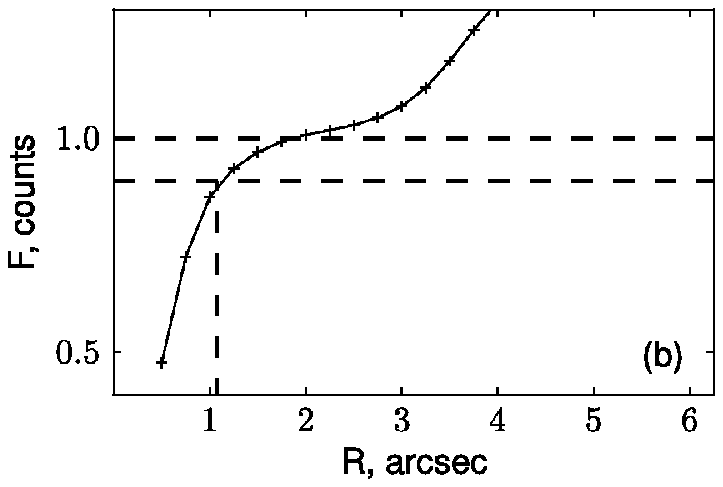}

\caption{Examples of typical variations of normalized stellar fluxes as a function of
 the aperture radius in two cases: the photometry of an isolated 
object (=) and the photometry of an object with a close neighbor
 (b). The crosses correspond to the results of measurements. The
 horizontal lines are drawn at the levels corresponding to 0.9 and
 1.0 total flux. The vertical line indicates the radius of the
 aperture corresponding to 0.9 of the total flux. In panel (b) the
 flux continues to increase after reaching the plateau because of
 the contribution from the neighboring object.}
\label{Flux000119}
\end{figure}

To choose the optimum aperture size, we performed the photometry
of each object varying the aperture radius R from 0\farcs{}5 to
6\farcs{}0 with a step of 0\farcs{}25. We estimated the background
level using the median averaging over a ring with the inner and
outer radii equal to R and R+$\Delta$R, respectively, where
\
{$\Delta {\rm R}=2''$}. In this approach the flux from a
single object first smoothly increases with increasing radius R
and then remains approximately constant (the plateau in
Fig.\,\ref{Flux000119}a). The vertical coordinate of the
point where the plateau begins corresponds to the full flux from
the star in the filter considered. After reaching the plateau the
flux becomes practically ceases to vary with aperture radius,
whereas the dependence is very strong when the radius approaches
the plateau. We consider the optimum radius of the object to be
equal to R$_{0.9}$,  the abscissa of the point corresponding to
90\% of the total flux. We interpolate the curve by drawing a
cubic spline through the measured data points.

In the presence of a nearby neighboring object (see  Fig.~\ref{Flux000119}b)
the flux continues to increase even after reaching the plateau. We chose the flux and optimum
aperture radius to be same for all filters at the supposed location of the plateau so as to
keep the flux from the neighboring objects outside the measurement aperture.

In the case of a two-dimensional Gaussian 90\% of the total flux
corresponds to R$_{0.9}=1.95\sigma$, where $\sigma$ is the
parameter of the Gaussian. The full width at half maximum (FWHM)
of a two-dimensional Gaussian (which, by definition, corresponds
to the size of a point source) is reached at the aperture radius
of R$_{FWHM}=1.18\sigma$. Hence given the observed
\mbox{0\farcs{}6--0\farcs{}8} seeing, we can expect the sizes of
point-source images on the CCD to be, according to our definition,
R$_{0.9}=1\farcs{}0-1\farcs{}3$.

To determine the optimum aperture R (or the size of the object),
we compute the mean R$_{0.9}$ values averaged over the four
broad-band filters. We then measure the flux in the narrow-band
\Ha{} filter within the same R$_{0.9}$ aperture and divide it by
0.9 to compute the total flux. The resulting quantity is equal to
the total flux if we are dealing with a point source in the \Ha{}
filter. In the case of an extended object in the \Ha{} filter
(e.g., an HII region), the inferred quantity is an estimate of the
total flux from the part of the extended source, where the point
source was measured in the broad-band filters.

Hence we use the R$_{0.9}$ quantity to measure the size of the source and use the count at the
plateau level to estimate the fluxes in filters as shown in Fig. \ref{Flux000119}.

The image of the galaxy consists of several zones with each zone
observed several times with different offsets so that many objects
are measured repeatedly in each particular filter. We corrected
each measurement for airmass by multiplying it by  $1/\cos{z}$,
where $z$ is the zenith angle. To construct the mean dependence of
flux on the aperture radius, we computed the median of the
dependences based on individual measurements. Individual
dependences differ mostly because of night-to-night seeing
variations. Averaging showed that the results of individual
measurements of the same object do not differ by more than
\mbox{7--10\%}, which is acceptable accuracy for our purposes of
object selection.

Massey et al.~[\cite{Massey2006}] demonstrated that each CCD
frame of the mosaic image should be reduced separately because of
considerable differences between the detector parameters (gain,
read-out noise, and spectral sensitivity). We therefore
partitioned each composite image into eight individual frames
corresponding to different detectors and reduced each frame
separately.

Our task of extracting objects with excess flux in the \Ha{} filter does not require the
determination of calibrated fluxes, and we therefore selected objects based on instrumental
fluxes exclusively. Only at the final stage of our work we converted instrumental fluxes
into magnitudes using the following averaged calibrating coefficients:

$$m_U = -2.5\log( F_U ) + 30.20$$
$$m_B = -2.5\log( F_B ) + 29.43$$
$$m_V = -2.5\log( F_V ) + 29.61$$
$$m_R = -2.5\log( F_R ) + 29.84$$
$$m_{{\rm H}\alpha} = -2.5\log( 2\times{}10^{-16} \times{} F_{HA} / 300 ),$$
where $F$ is the flux from the object measured during the exposure time in the instrumental
system in the corresponding filter. We determined the averaged calibrating coefficients by
performing the photometry of single stars and comparing their instrumental magnitudes
with the magnitudes reported in the catalog of Massey et al.~[\cite{Massey2006}], and
only adopted the \Ha{} calibration from~[\cite{Massey2007_Ha}]. Note once again that
our main task requires no flux calibration.

\section{SELECTION OF LBV-LIKE CANDIDATE OBJECTS}

For our photometric measurements we selected a total of 2304
objects with $V <$\Vlim{} from the catalog of Massey et
al.~[\cite{Massey2006}] . We adopt an average interstellar
extinction of A$_V$$\approx$1\fm0 for the brightest stars in the
M\,33 galaxy (e.g.,~[\cite{Fabrika_etal1999}]
A$_V$$\approx$0\fm95$\pm$0\fm05), and an M\,33 distance modulus of
24\fm9~[\cite{M33Distance}]. Given our adopted criteria
($V <$\Vlim{} and  $(B-V) <$\BVlim{}) the main list includes stars
with M$_V < -$7\fm4 and \mbox{$(B-V)_0\le$0\fm0}. It includes all
bright supergiants of liminosity class Iab and more luminous
stars, as well as the hottest supergiants (with B0-type and
earlier spectra) of luminosity class Ib [\cite{Straizhis}].
Hence our final list should include all potential LBV star
candidates.

In the case of the photometry of close groups consisting of two or
more objects, where the plateau in the ``aperture radius--flux''
dependence for all broad-band filter images can be reached at
apertures greater than the aperture of a point source, we treated
such a close group as a single object. It is evident that some of
such ``single'' objects will prove to be groups of stars if
observed with a better spatial resolution. In our photometric
measurements a close group could consist of several objects from
the initial catalog of Massey et al.~[\cite{Massey2006}]. We
attributed the resulting measurement to one of the objects of the
initial catalog and did not perform the photometry of other
members of the group. If one of the objects in such a close group
is an  \Ha{} emission star the entire group shows excess flux in
the \Ha{} filter.

We were able to perform the photometry of 2075 objects.
Photometric measurements were impossible to perform for the
remaining 229 objects, because they were  located at the boundary
of the galaxy, showed no plateau on the ``aperture radius--flux''
dependence, or were members of one of the close groups.

\begin{figure*}
\includegraphics{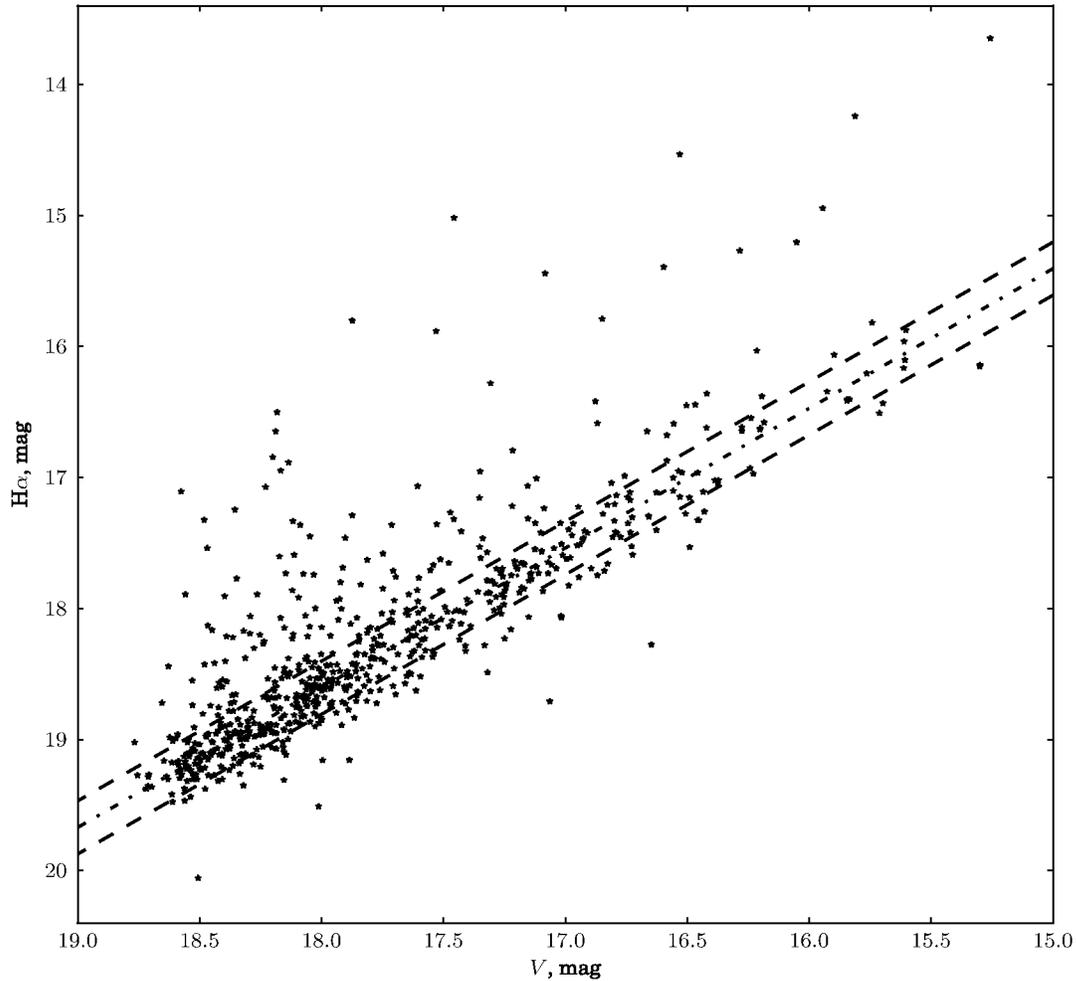}
 \caption{\Ha{}-band versus $V$-band
magnitude diagram for 707 objects with $B-V<$\BVlim{}. The
dashed-and-dotted line shows the relation for the nonemission
sample. The dotted lines show the  $\pm1\sigma$ levels. Objects
located above 1$\sigma$ level are listed in
Table~\ref{FullList} as emission-line objects.}
\label{selectVHA}
\end{figure*}

\begin{figure*}
\includegraphics{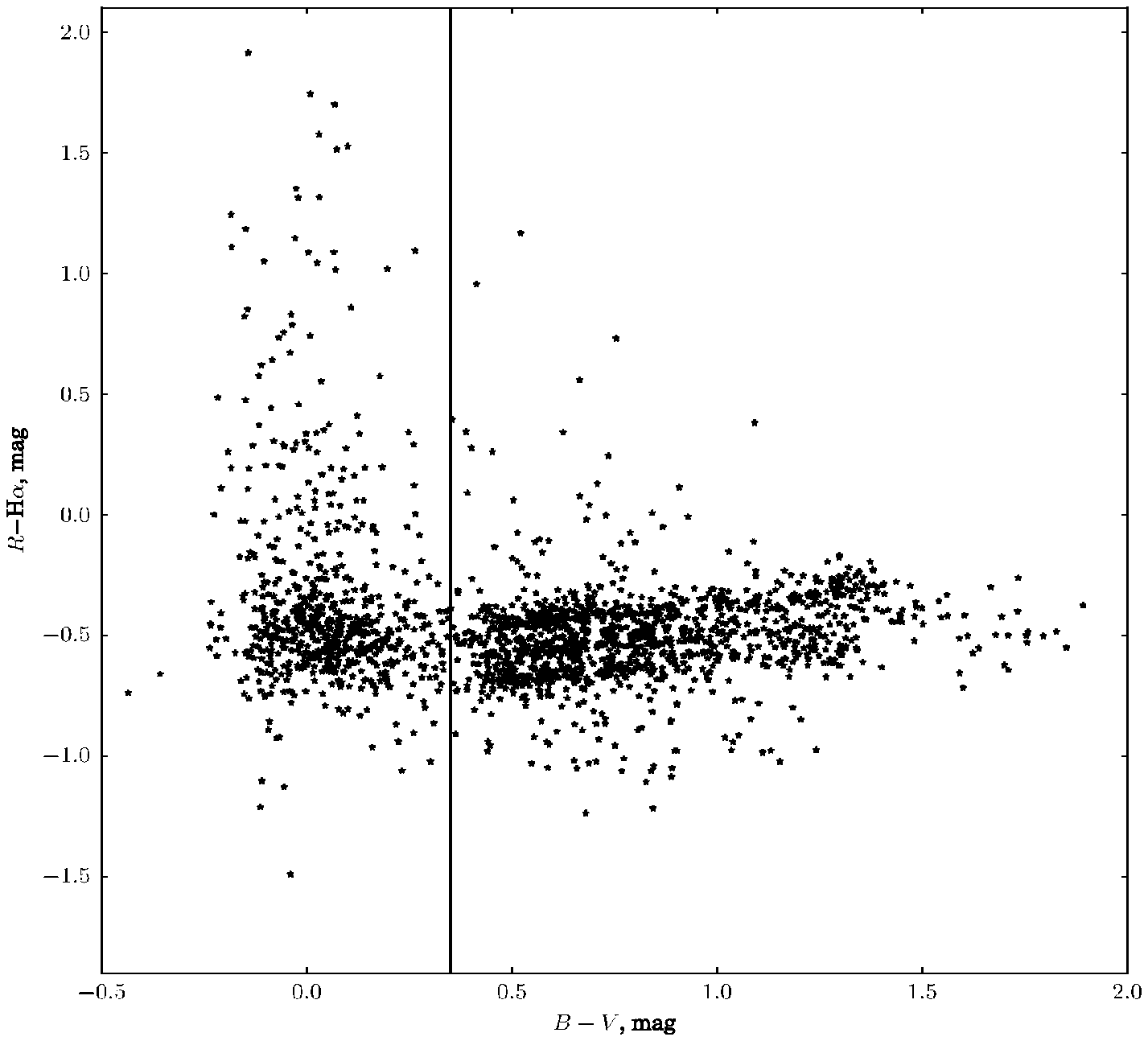}
 \caption{Results of the photometry of all
the 2075 sample objects with $V <$\Vlim{}. Objects with
$B-V <$\BVlim{} in the top left part of the figure were
investigated based on the criterion illustrated in
Fig.~\ref{selectVHA}. Objects in the top right part of the
figure with an $R-$\Ha{} color excess are listed in
Table~\ref{FullListRed} as potential LBV candidates with
strong interstellar extinction. } \label{selectB_VR_HA}
\end{figure*}

After the photometric reduction we selected the emission-line candidates using the method
described by Fabrika et al.~[\cite{Fabrika_etal1997}]. To this end, we constructed the
$V-$\Ha{} diagram, where most of the objects lie within a broad band of a linear relation
(Fig. \ref{selectVHA}). Emission-line objects lie above this relation because of the
appreciable  excess of their \Ha{} fluxes relative to the remaining, nonemission part of the
sample. Figure~\ref{selectVHA} shows an example of the $V-$\Ha{} relation.

It is evident that this linear  ``main sequence'' is due to the
fact that stars that are brighter in the V band are also
relatively brighter in the \Ha{} band. This ``main sequence'' is
rather broad due both to absorption stars and to spectral
peculiarities of stars of different temperatures. However, it is a
well-defined sequence.

At the next stage we used only 707 stars with $B-V<$\BVlim{} out
of the entire sample of 2075 objects. We show these 707 stars in
Fig.~\ref{selectVHA}. The  $V-$\Ha{} diagram for the entire
sample including both red and blue stars would consist of two
partially overlapping parallel sequences: the red stars in  M\,33
and in Milky Way.

We used the following technique to select emission-line stars on the $V-$\Ha{} diagram. We first
fitted a linear relation to all points of the sample in order to determine the main sequence of
nonemission stars, and then computed the r.m.s. deviation (hereafter referred to as $\sigma{}$)
of data points from this linear relation. At the next stage we discarded the objects with the
\Ha{} excess greater than $2.5\sigma{}$; fitted a new relation, and
computed the new $\sigma{}$ value based on the remaining data points.
We computed the final $\sigma{}$ for the rectified sequence obtained after seven to eight
iterations when the number of stars ceased to change.

We then choose the minimum excess such that objects lying above it
should be considered to be emission-line stars. We found that the
\Ha{}7 object with an \Ha{} emission-line equivalent width of
3\AA{} [\cite{Shol_etal97}] lies $1.01\sigma{}$ above the
main relation (number 018246). We therefore decided that emission
objects should lie about $1.0\sigma{}$ above the main sequence and
used this criterion to select 185 objects, which we list in
Table~\ref{FullList}.

The first three columns of Table~\ref{FullList} list the
number of the object and its J2000.0 coordinates according to the
catalog of Massey et al.~[\cite{Massey2006}]. The next
columns give our measured $V$-band magnitude and the $(B-V)$ color
index; the excess ``s'' above the main sequence in the units of
$\sigma$ (used as the estimate of the intensity of the
hypothetical \Ha{} emission), and R, the aperture size R$_{0.9}$
averaged over all four broad-band filters (in arcseconds). The
complete version of the table is available at {\tt
http://jet.sao.ru/$\sim$azamat/LBVsearch/\linebreak /blue.dat}.

Given that red stars with $(B-V)\ge$\BVlim{} can also be LBV
candidates (e.g., the highly reddened $\eta$\,Car), we selected
from our list the candidate emission-line objects among other
stars  with $V <$\Vlim{}. We then used the method described above
to compute the r.m.s. deviation $\sigma$ of the  $(B-V)-(R-$\Ha{})
diagram for objects with $(B-V)\ge$\BVlim{} shown in
Fig.~\ref{selectB_VR_HA}. We included into the list of LBV
candidates (Table~\ref{FullListRed}) only the objects with
the excess of the ($R-$\Ha{}) color excess greater than 2$\sigma$
above the main sequence of stars with  $(B-V)\ge$\BVlim{}.

We imposed an additional constraint \linebreak $(B-V) <$1\fm2 on
the colors of red stars in Table~\ref{FullListRed}. Note
that no objects with \Ha{} excess were selected in the  $(B-V) =
1\fm2 - 1\fm3$ interval. Given the known properties of LBV
stars~[\cite{HD1994_LBVreview}] it can be concluded that
their photospheric temperatures cannot be lower than 7000K, which
corresponds to $(B-V)_0\approx$0\fm25. The observed color of such
a star should be $(B-V)\approx$1\fm2 even if the interstellar
extinction toward it were as high as A$_V=3\fm0$. Moreover, the
spectra of stars with  $(B-V)_0>$1\fm3 exhibit TiO absorption
bands, which may mimic  \Ha{} emission. In view of these
considerations, our constraint  $(B-V) <$1\fm2 appears to be quite
reasonable.

\begin{figure}
\includegraphics{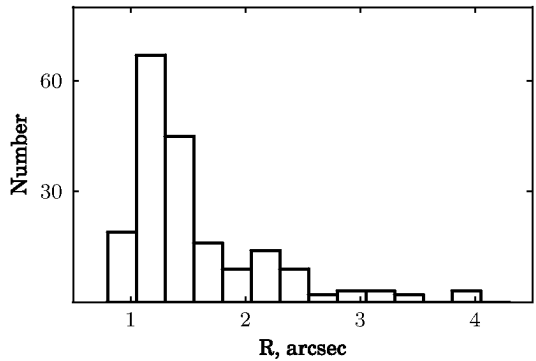}
 \caption{Histogram of the R$_{0.9}$ sizes of our selected
objects (Table~\ref{FullList}). The maximum of this distribution, which is located at
1\farcs{}0-1\farcs{}3, corresponds to the expected size of point sources (see Section
\ref{Sect03}). Sources with greater sizes correspond to groups, which we treated as
single objects in our photometric measurements.} \label{sizeHist}
\end{figure}

\begin{figure}
\includegraphics{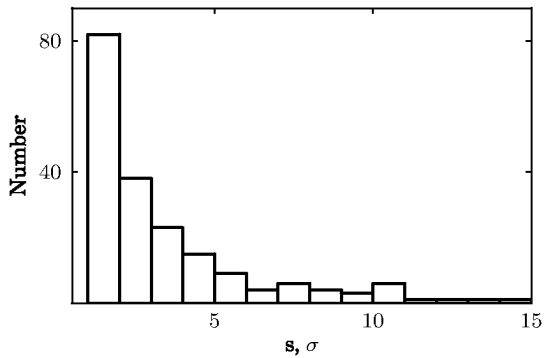}
 \caption{Histogram of the \Ha{} flux excess above the main sequence for
nonemission stars in the units of $\sigma$ (Table~\ref{FullList}).}
\label{sigmaHist}
\end{figure}

\section{COMPARISON WITH OTHER CATALOGS}
We cross-correlated the coordinates of objects of our list of blue
candidate emission-line stars with other published lists. To this
end, we used TOPCAT\footnote{TOPCAT was initially (2003-2005)
developed under the UK Starlink project (1980-2005, R.I.P.). From
July 2005 until June 2006, it was supported by grant PP/D002486/1
from the UK's Particle Physics and Astronomy Research Council.
Maintenance and development has been funded from July 2006 until
December 2007 by the European VOTech project within the UK's
AstroGrid, and directly from AstroGrid funding beyond that.}
program, running the standard identification algorithm where
objects from the first catalog are selected that are located
inside the $\varepsilon$-neighhborhood of an object of the second
catalog. We set the radius of $\varepsilon$-neighborhood (or the
``error box'') individually for each pair of catalogs.

The cross-correlation with the lists of sources detected by
Chandra [\cite{M33ChandraList}] and XMM Newton
[\cite{M33XMMList}] X-ray telescopes with 5\farcs{}0-large
error boxes yielded one identification. This X-ray object (the
supernova remnant SNR B013059+30177~[\cite{M33SNRList}]) has
already certain optical identification and is located near the
N085197 (J013348.03+303304.8) object from
Table~\ref{FullList}. A cross-correlation with the list of
ultraviolet sources [\cite{MasseyUIT}](UIT) yielded 23
identifications within 3\farcs{}0-large error boxes, of which only
four objects have not yet been spectrally classified.

Classical LBV stars are characterized by irregular light
variations on various time scales, and therefore we
cross-correlated our list with the catalog of variable stars in
M\,33 [\cite{M33VariableList}], which contains more than
36000 variables down to a limiting magnitude of
\mbox{$V\approx24$\fm0} with the coordinates accurate  to
0\farcs{}2. We found 54 stars in common,
which makes up for  29\% of our complete list of blue objects with  \Ha{} emission.

We also cross-correlated our list with two lists of emission-line
objects composed in different \linebreak
years~[\cite{Massey2007_Ha}], [\cite{Fabrika_etal1997}].
The evident criterion based on the exact coincidence of the
coordinates in our list with those of LBV candidates selected by
Massey et al.~[\cite{Massey2007_Ha}] yielded 113 stars in
common, whereas the cross-correlation with the catalog of Fabrika
et al.~[\cite{Fabrika_etal1997}] with a 2\farcs{}0-large
error box yielded 50 identifications.

\section{DISCUSSION}

Figure~\ref{sizeHist} shows the size distribution of 185 selected blue emission-line
objects from Table~\ref{FullList}. The maximum of the distribution is near
1\farcs{}0--1\farcs{}3, which corresponds to  the value expected for point objects observed with
a seeing of 0\farcs{}6--0\farcs{}8 (see Section~\ref{Sect03}). Sources with greater sizes
correspond to groups of objects.

Figure~\ref{sigmaHist} shows the distribution of the flux
excess ``s'' in the \Ha{} filter for selected blue objects above
the main linear relation for nonemission objects. As is evident
from the figure, the number of objects continuously decreases with
increasing  \Ha{}-line flux excess.

Note that Table~\ref{FullList} contains no objects in the
$V\approx14$\fm0--15\fm5 magnitude interval. This means that seven
objects with V$<$14\fm0 must be foreground stars of the Milky Way,
although, only spectroscopic observations can explain the excess
of \Ha{}-line fluxes of these objects. A 62\% overlap between our
list of blue objects and the list of emission-line objects
of~[\cite{Massey2007_Ha}] shows that both criteria can be
used to select LBV candidates. Non cross-identified objects of
both lists also deserve a special investigation in order to
understand what subclass of emission-line objects is rejected by
each method.

We included into our additional list of LBV candidates
(Table~\ref{FullListRed}) only objects with $V>16$\fm{}0,
because no red objects were selected in the $V\approx
15$\fm0--16\fm0 magnitude interval, and brighter stars are
definitively foreground objects. The fact that two red emission
objects (N006862 and N141751) are also included in the catalog of
Massey et al.~[\cite{Massey2007_Ha}], where they are
classified as hot LBV candidates, corroborates the need for the
search for reddened stars.

We perform spectroscopic observations of blue emission objects
with the 6-m telescope of the Special Astrophysical Observatory of
the Russian Academy of Sciences. We already
found~[\cite{NewLBVinM33}] the star N93351 whose spectrum
exhibits broad hydrogen emissions and numerous FeII and [FeII]
emission features. We constructed the spectral energy distribution
for this star in the wavelength interval \linebreak
3000--80000\AA{} and showed that this object, like Var\,A, has a
strong infrared excess. The results of our analysis led us to
conclude~[\cite{NewLBVinM33}] that N93351 should be
classified as an LBV-type star.

\section{CONCLUSIONS}

We used archive of UBVR and \Ha{}-band CCD images of the M\,33
galaxy to perform aperture photometry of all objects from
catalog~[\cite{Massey2006}] with $V <$\Vlim{}. We selected LBV
candidate stars in M\,33 using a criterion based on the excess of
the \Ha{}-line flux over the $V$-band flux. We selected a total of
185 emission-line candidates --- blue stars with $(B-V) <$\BVlim{}.

Variability on different time scales is a classical property of
LBV stars and therefore we cross-correlated our list with the
catalog of variable stars in  M\,33~[\cite{M33VariableList}]
to find that 29\% of our blue candidates exhibit light variations.
Part of LBV stars are hot and should be studied in the
ultraviolet. A cross-correlation with the list of ultraviolet
sources showed that 23 LBV candidates were indeed detected by the
UIT space telescope [\cite{MasseyUIT}]. Our list of blue
emission-line candidates overlaps by 27\% with the list of Fabrika
et al.~[\cite{Fabrika_etal1997}] based on photometric images
and simple criteria. Our list also overlaps by  62\% with the list
of LBV candidates~[\cite{Massey2007_Ha}] based on the same
observational material, but subject to more complex criteria of
the selection of LBV candidates.

LBV stars may be appreciably, and sometimes rather strongly, reddened. We therefore composed an
additional list of 25 red stars with \BVlim{}$< B-V <$1\fm{}2 selected using the criterion of the
excess of the  $R-$\Ha{} color.  These objects may be reddened LBV stars.

We thus report our complete list of LBV candidates in  M\,33 down
to a $V$-band limiting magnitude of \Vlim{}. To perform a detailed
classification of the stars of our catalog, we observe them
spectroscopically with the 6-m telescope of the Special
Astrophysical Observatory of the Russian Academy of Sciences. We
will publish the results of these spectroscopic observations in
our next papers.

\section*{acknowledgments}
This work was supported by the Russian Foundation for Basic
Research (project nos\,09-02-00163 and~07-02-00909).

\bibliography{allbibl}
\bibliographystyle{mn2e}

\newpage
\onecolumn
\begin{longtable}{c|c|c|c|c|c|c|l}
 \caption{LBV candidates with $(B-V) <$\BVlim{}. The columns give
 (in this order): the number of the object and its coordinates
%   adopted from~[\cite{Massey2006}]; our measured $V$-band
   adopted from~[Massey et~al. (2006)]; our measured $V$-band
 magnitude and $(B-V)$ color index;  the excess ``s'' of the
 \Ha-line flux above main sequence in the units of  $\sigma$, and
 R, the object's size R$_{0.9}$ averaged over four broad- band
 filters in arcseconds. The last column gives the results of
 cross-identification with the catalogs of emission-line objects
%  ([\cite{Massey2007_Ha}] (M) and
%  [\cite{Fabrika_etal1997}] (SFZ)); variable
%  stars~[\cite{M33VariableList}] (HBS), and ultraviolet
%  sources~[\cite{MasseyUIT}] (UIT). If the object has a comment
%  in catalog~[\cite{Massey2006}] we repeat this comment here in
%  the ``m(comment)'' format. The full version of the table is
%  available from 
%  {\tt http://jet.sao.ru/$\sim$azamat/LBVsearch/blue.dat}.
 ([Massey et~al. (2007)] (M) and
 [Fabrika et~al. (1997)] (SFZ)); variable
 stars~[Hartman et~al. (2006)] (HBS), and ultraviolet
 sources~[Massey et~al. (1996)] (UIT). If the object has a comment
 in catalog~[Massey et~al. (2006)] we repeat this comment here in
 the ``m(comment)'' format. The full version of the table is
 available from 
 {\tt http://jet.sao.ru/$\sim$azamat/LBVsearch/blue.dat}.
 }
\label{FullList}\\

\hline%\hline
N & $\alpha$ &  $\delta$ & V & $B-V$ & s &R($''$) & comment\\
\hline
1 & 2 & 3 & 4 & 5 & 6 & 7 & 8 \\
\hline
\endfirsthead
\caption{(Contd.)}\\

\hline
1 & 2 & 3 & 4 & 5 & 6 & 7 & 8 \\
\hline
\endhead

\hline
\endfoot

\hline%\hline
\endlastfoot

000029 & 01:31:48.14 & 30:32:06.9 & 17.61 & $-$1.10 & 2.04 & 1.18 & \\ 
001429 & 01:32:27.81 & 30:21:46.8 & 18.40 & 0.32 & 3.09 & 0.95 & \\ 
001705 & 01:32:29.08 & 30:34:04.2 & 17.71 & 0.30 & 1.73 & 2.23 & SFZ515, HBS260273\\ 
003562 & 01:32:34.31 & 30:38:17.3 & 13.58 & $-$0.21 & 2.41 & 1.1 & \\ 
003935 & 01:32:35.25 & 30:30:17.6 & 18.08 & 0.04 & 4.69 & 1.98 & HBS250024, M(hot LBV cand)\\ 
004926 & 01:32:37.72 & 30:40:05.6 & 17.48 & $-$0.03 & 3.83 & 0.93 & SFZ005, UIT003, m(Ofpe/WN9\_M33WR2), M\\ 
005705 & 01:32:39.64 & 30:24:51.9 & 18.30 & $-$0.09 & 4.24 & 2.15 & M\\ 
006389 & 01:32:41.30 & 30:22:31.2 & 18.10 & $-$0.03 & 3.9 & 1.32 & HBS340115, M(HII)\\ 
008043 & 01:32:44.62 & 30:34:59.5 & 18.14 & $-$0.15 & 9.21 & 1.23 & HBS250385, M\\ 
008581 & 01:32:45.68 & 30:39:06.9 & 18.19 & $-$0.14 & 11.35 & 0.82 & HBS250451\\ 
008632 & 01:32:45.75 & 30:38:55.8 & 16.88 & $-$0.04 & 4.89 & 1.4 & m(WN\_M33WR6)\\ 
009925 & 01:32:48.26 & 30:39:50.4 & 17.22 & 0.09 & 4.82 & 2.48 & SFZ517, M\\ 
012779 & 01:32:52.95 & 30:34:50.3 & 18.10 & 0.07 & 1.18 & 1.18 & \\ 
013846 & 01:32:54.37 & 30:30:50.6 & 17.88 & 0.24 & 1.78 & 1.07 & \\ 
014939 & 01:32:55.68 & 30:35:34.7 & 17.65 & 0.04 & 1.64 & 1.15 & SFZ022,m(B5Ia\_ob21$-$40)\\ 
015651 & 01:32:56.47 & 30:35:30.9 & 18.53 & $-$0.04 & 1.26 & 1.73 & HBS250999\\ 
017442 & 01:32:58.71 & 30:31:52.9 & 17.96 & 0.07 & 1.05 & 1.05 & HBS251139\\ 
018246 & 01:32:59.74 & 30:38:54.8 & 17.77 & 0.14 & 1.01 & 0.88 & SFx xZ033, M(HII?)\\ 
019512 & 01:33:01.24 & 30:30:51.3 & 18.30 & 0.22 & 3.59 & 1.05 & M(HII 2)\\ 
021012 & 01:33:03.09 & 30:31:01.8 & 17.16 & 0.09 & 3.16 & 1.18 & SFZ041, M(HII)\\ 
021189 & 01:33:03.31 & 30:11:21.8 & 18.20 & $-$0.07 & 1.77 & 1.25 & M\\ 
021331 & 01:33:03.50 & 30:33:23.1 & 18.39 & $-$0.13 & 4.0 & 2.0 & SFZ042, HBS251493,M\\ 
024824 & 01:33:07.60 & 30:42:59.0 & 16.76 & 0.34 & 1.44 & 1.85 & SFZ049\\ 
024835 & 01:33:07.61 & 30:42:43.0 & 18.41 & $-$0.14 & 2.44 & 1.23 & M\\ 
025221 & 01:33:08.04 & 30:46:12.8 & 12.98 & 0.08 & 1.09 & 1.65 & \\ 
025981 & 01:33:08.99 & 30:29:56.3 & 17.53 & 0.14 & 3.68 & 1.55 & m(B8Iacomp\_ob127$-$15),M\\ 
026091 & 01:33:09.14 & 30:49:54.5 & 18.24 & $-$0.01 & 2.92 & 0.95 & SFZ052, HBS150200, m(Ofpe/WN9\_M33WR22,UIT045),M\\ 
027321 & 01:33:10.43 & 30:38:49.4 & 17.85 & 0.10 & 1.17 & 1.32 & M \\ 
028115 & 01:33:11.25 & 30:45:15.4 & 15.26 & 0.10 & 10.04 & 3.92 & M, UIT049, m(BI+neb) \\ 
028158 & 01:33:11.30 & 30:29:33.5 & 18.18 & 0.18 & 5.86 & 3.92 & HBS240638 \\ 
028322 & 01:33:11.45 & 30:29:51.3 & 17.12 & 0.06 & 3.25 & 2.2 & M(HII)\\ 
028576 & 01:33:11.70 & 30:22:58.9 & 17.75 & 0.28 & 2.42 & 1.43 & \\ 
028741 & 01:33:11.85 & 30:38:52.7 & 16.59 & $-$0.01 & 2.06 & 1.25 & \\ 
028771 & 01:33:11.88 & 30:38:53.5 & 16.42 & 0.00 & 2.77 & 1.57 & m(WNL\_M33WR25=MC17)\\ 
029660 & 01:33:12.81 & 30:30:12.6 & 17.48 & 0.17 & 1.96 & 1.07 & HBS240772,M \\ 
030009 & 01:33:13.17 & 31:04:59.3 & 18.20 & 0.08 & 1.01 & 0.95 & \\ 
031983 & 01:33:15.21 & 30:37:27.2 & 18.36 & 0.15 & 1.61 & 1.27 & HBS241026\\ 
032127 & 01:33:15.34 & 30:37:22.6 & 17.13 & 0.34 & 1.59 & 1.77 & \\ 
032629 & 01:33:15.82 & 30:56:44.8 & 16.51 & $-$0.00 & 2.76 & 1.57 & m(WN4.5+neb\_M33WR33=MC23),M\\ 
033347 & 01:33:16.50 & 30:32:12.1 & 17.09 & 0.26 & 1.96 & 1.4 & SFZ080, M(late A~I/early F~I) \\ 
033824 & 01:33:16.92 & 30:23:07.3 & 18.49 & $-$0.00 & 1.59 & 1.38 & SFZ084\\ 
037649 & 01:33:20.58 & 30:31:42.4 & 18.61 & 0.10 & 1.21 & 1.15 & \\ 
045716 & 01:33:27.26 & 30:39:09.1 & 18.04 & $-$0.12 & 4.43 & 1.27 & HBS242552,m(Ofpe/WN9\_M33WR39=MJC7),M \\ 
045922 & 01:33:27.41 & 30:31:31.7 & 18.38 & $-$0.13 & 1.13 & 2.2 & SFZ107,HBS242580,M\\ 
049612 & 01:33:29.88 & 30:31:47.3 & 17.88 & 0.01 & 13.17 & 1.27 & M,UIT113,m(O+neb) \\ 
051296 & 01:33:30.99 & 30:36:52.5 & 18.37 & 0.22 & 1.65 & 1.23 & SFZ119,HBS243138 \\ 
053273 & 01:33:32.23 & 30:41:31.2 & 18.35 & $-$0.15 & 5.95 & 2.3 & M\\ 
054233 & 01:33:32.82 & 30:41:46.0 & 18.17 & $-$0.19 & 3.52 & 1.27 & m(WNL\_M33WR42=AM3),M\\ 
054460 & 01:33:32.97 & 30:41:36.1 & 18.27 & $-$0.15 & 4.91 & 1.05 & m(WNL\_M33WR43=AM4),M\\ 
054488 & 01:33:32.98 & 30:33:44.2 & 18.53 & 0.16 & 3.07 & 1.35 & HBS243502\\ 
054767 & 01:33:33.13 & 30:35:06.3 & 17.72 & $-$0.01 & 1.06 & 1.07 & SFZ124,m(B8I\_R93$-$8),M\\ 
056807 & 01:33:34.27 & 30:41:36.7 & 16.54 & 0.07 & 12.37 & 2.08 & M(WNL in HII)\\ 
056847 & 01:33:34.29 & 30:34:00.1 & 17.86 & 0.13 & 1.88 & 1.35 & M(HII)\\ 
057017 & 01:33:34.39 & 30:32:08.4 & 18.20 & $-$0.02 & 9.75 & 1.35 & M(HII)\\ 
058382 & 01:33:35.14 & 30:36:00.4 & 16.67 & 0.12 & 2.64 & 1.3 & m(LBV\_VarC),M \\ 
058864 & 01:33:35.40 & 30:35:54.7 & 18.66 & 0.07 & 2.88 & 2.15 & \\ 
060182 & 01:33:36.15 & 30:50:37.2 & 18.07 & $-$0.12 & 1.44 & 1.57 & UIT138,m(O6.5II),M\\ 
060689 & 01:33:36.42 & 30:35:30.9 & 18.03 & 0.05 & 1.11 & 1.27 & SFZ141,HBS230123\\ 
060906 & 01:33:36.54 & 30:20:58.2 & 18.38 & 0.03 & 1.39 & 1.38 & HBS320012,M\\ 
062775 & 01:33:37.58 & 30:28:04.7 & 18.15 & 0.16 & 1.27 & 2.27 & \\ 
063555 & 01:33:38.01 & 30:31:48.6 & 18.06 & 0.07 & 2.24 & 2.12 & SFZ149,HBS230517,M\\ 
065919 & 01:33:39.24 & 30:20:22.5 & 18.01 & 0.09 & 1.28 & 1.4 & HBS320142\\ 
065935 & 01:33:39.25 & 30:43:03.6 & 18.63 & $-$0.14 & 1.42 & 1.1 & HBS140447,M\\ 
066011 & 01:33:39.28 & 30:20:53.6 & 18.10 & $-$0.09 & 1.35 & 1.2 & HBS320143,M\\ 
066512 & 01:33:39.52 & 30:45:40.5 & 17.31 & 0.11 & 7.83 & 4.42 & SFZ163,m(MB0.5Ia+WNE\_M33WR57),UIT154,M(P Cyg LBV cand)\\ 
069329 & 01:33:40.82 & 30:31:32.6 & 17.70 & 0.02 & 2.58 & 2.02 & SFZ186,HBS231250,M \\ 
069374 & 01:33:40.84 & 30:38:22.5 & 18.17 & 0.18 & 1.22 & 1.52 & HBS231309\\ 
070474 & 01:33:41.33 & 30:32:12.6 & 18.41 & $-$0.00 & 2.19 & 1.07 & SFZ191,m(O8:If\_ob10$-$3),M\\ 
071501 & 01:33:41.80 & 30:21:10.7 & 18.03 & 0.14 & 3.13 & 2.45 & HBS320276\\ 
072150 & 01:33:42.08 & 30:42:00.3 & 18.37 & 0.02 & 3.82 & 1.25 & HBS140882,M\\ 
073722 & 01:33:42.78 & 30:32:56.3 & 18.12 & $-$0.04 & 6.9 & 1.32 & SFZ207,M \\ 
074495 & 01:33:43.10 & 30:39:04.5 & 18.05 & $-$1.07 & 2.57 & 1.02 & \\ 
074593 & 01:33:43.14 & 30:39:07.6 & 16.43 & 0.02 & 1.49 & 2.05 & \\ 
075005 & 01:33:43.31 & 30:35:33.8 & 17.65 & 0.02 & 1.1 & 1.07 & \\ 
075058 & 01:33:43.34 & 30:35:34.1 & 17.65 & 0.02 & 1.1 & 1.07 & m(WN4.5+O6$-$9\_M33WR75=UIT177)\\ 
076123 & 01:33:43.82 & 30:32:10.2 & 17.55 & 0.00 & 2.22 & 1.75 & \\ 
076579 & 01:33:44.02 & 30:33:18.2 & 18.53 & $-$0.10 & 2.14 & 1.38 & M\\ 
076951 & 01:33:44.18 & 30:31:24.0 & 18.43 & $-$0.08 & 1.49 & 1.32 & SFZ217,M\\ 
077511 & 01:33:44.43 & 30:38:43.9 & 17.94 & 0.05 & 2.21 & 1.32 & SFZ219,M\\ 
077826 & 01:33:44.56 & 30:32:01.3 & 18.36 & 0.00 & 8.58 & 1.98 & SFZ221,M(HII)\\ 
078046 & 01:33:44.65 & 30:35:59.2 & 17.34 & $-$0.02 & 2.15 & 1.23 & m(B1.5Ia+\_W91$-$258),M\\ 
078106 & 01:33:44.68 & 30:44:36.7 & 18.19 & $-$0.03 & 10.65 & 1.38 & m(WN8\_M33WR77=OB66$-$25),M\\ 
078287 & 01:33:44.75 & 30:44:44.5 & 18.45 & 0.26 & 4.54 & 1.12 & M\\ 
078412 & 01:33:44.79 & 30:44:32.4 & 18.23 & 0.07 & 8.78 & 1.38 & HBS141343,M,m(OB+neb\_ob66$-$28s),M(HII)\\ 
078458 & 01:33:44.81 & 30:32:17.8 & 18.12 & $-$0.04 & 2.49 & 1.45 & SFZ222, HBS232168,M\\ 
078579 & 01:33:44.87 & 30:32:11.0 & 18.77 & $-$0.08 & 1.98 & 1.45 & \\ 
078781 & 01:33:44.97 & 30:36:16.9 & 17.56 & $-$0.08 & 2.08 & 1.3 & M\\ 
079431 & 01:33:45.25 & 30:36:26.6 & 17.75 & $-$0.08 & 3.73 & 1.23 & M(HII),m(BI\_W91$-$245)\\ 
080679 & 01:33:45.86 & 30:44:44.5 & 17.82 & 0.26 & 3.84 & 1.82 & \\ 
082991 & 01:33:46.96 & 30:36:42.8 & 18.16 & 0.24 & 1.54 & 1.05 & \\ 
083098 & 01:33:47.01 & 30:46:17.5 & 13.46 & 0.18 & 1.37 & 1.1 & \\ 
083744 & 01:33:47.33 & 30:33:06.8 & 17.94 & 0.02 & 2.96 & 1.23 & HBS232867,M(HII)\\ 
084795 & 01:33:47.82 & 30:43:24.9 & 17.72 & 0.10 & 1.08 & 3.17 & SFZ237, HBS141903\\ 
085197 & 01:33:48.03 & 30:33:04.8 & 17.53 & 0.03 & 10.97 & 3.83 & m(HII),UIT205\\ 
087513 & 01:33:49.22 & 30:38:09.3 & 16.29 & 0.03 & 7.44 & 1.18 & M\\ 
087530 & 01:33:49.23 & 30:38:09.1 & 16.29 & 0.03 & 7.44 & 1.18 & SFZ246, m(LBV\_VarB)\\ 
089263 & 01:33:50.12 & 30:41:26.6 & 16.60 & 0.20 & 8.47 & 1.25 & SFZ256,m(LBVcand),UIT212,M(LBVcand)\\ 
091640 & 01:33:51.42 & 30:40:00.7 & 18.39 & 0.17 & 2.29 & 1.3 & M\\ 
091701 & 01:33:51.46 & 30:40:57.0 & 17.36 & 0.09 & 1.9 & 1.48 & M(P Cyg LBV cand)\\ 
091715 & 01:33:51.47 & 30:38:48.7 & 18.13 & 0.01 & 1.25 & 1.1 & SFZ268,M\\ 
092302 & 01:33:51.80 & 30:38:49.0 & 18.24 & $-$0.14 & 2.99 & 1.35 & SFZ271,M\\ 
092768 & 01:33:52.07 & 30:52:50.7 & 18.22 & $-$0.08 & 1.51 & 1.3 & \\ 
092935 & 01:33:52.16 & 30:33:33.5 & 18.00 & 0.16 & 2.28 & 1.65 & HBS234175,M\\ 
092983 & 01:33:52.19 & 30:36:36.6 & 17.92 & 0.25 & 2.55 & 1.27 & M(HII)\\ 
093303 & 01:33:52.39 & 30:39:20.9 & 16.85 & $-$0.10 & 7.85 & 1.38 & M(HII),m(OB+neb),UIT229\\ 
093351 & 01:33:52.42 & 30:39:09.6 & 16.20 & 0.24 & 1.48 & 1.32 & new LBV [\cite{NewLBVinM33},M\\ 
093765 & 01:33:52.66 & 30:39:13.9 & 17.36 & $-$0.05 & 3.75 & 1.6 & m(B5:I\_UIT231),M\\ 
094256 & 01:33:52.95 & 30:44:57.0 & 15.75 & 0.30 & 1.87 & 1.25 & \\ 
094642 & 01:33:53.21 & 30:38:54.0 & 17.22 & 0.05 & 2.73 & 1.35 & \\ 
095270 & 01:33:53.60 & 30:38:51.6 & 18.05 & $-$0.12 & 5.96 & 1.35 & HBS234654,m(Ofpe/WN9\_M33WR103=MJX15),M\\ 
095281 & 01:33:53.61 & 30:38:43.0 & 18.47 & 0.25 & 4.81 & 1.35 & \\ 
096035 & 01:33:54.10 & 30:33:09.7 & 17.09 & 0.07 & 10.79 & 1.98 & m(O6$-$8If),UIT240\\ 
096860 & 01:33:54.64 & 30:33:08.2 & 17.71 & 0.06 & 2.86 & 1.52 & \\ 
097162 & 01:33:54.84 & 30:32:48.9 & 18.48 & $-$0.06 & 3.41 & 1.32 & SFZ294,HBS234917,M \\ 
097751 & 01:33:55.21 & 30:34:29.9 & 17.16 & 0.03 & 1.92 & 2.38 & m(B5Ia\_ob4$-$4),M\\ 
098246 & 01:33:55.51 & 30:45:26.8 & 17.93 & 0.01 & 3.57 & 1.18 & M(HII)\\ 
098632 & 01:33:55.75 & 30:45:30.1 & 18.44 & $-$0.18 & 3.25 & 0.93 & M\\ 
098810 & 01:33:55.87 & 30:45:28.4 & 17.32 & $-$0.23 & 1.54 & 0.97 & M(WNL in HII)\\ 
100400 & 01:33:56.85 & 30:34:29.7 & 17.81 & $-$0.16 & 1.21 & 2.9 & \\ 
100647 & 01:33:57.00 & 30:38:26.4 & 17.61 & $-$0.15 & 1.61 & 2.85 & SFZ301,M\\ 
101408 & 01:33:57.45 & 30:32:27.6 & 18.26 & 0.08 & 3.33 & 2.12 & \\ 
101826 & 01:33:57.69 & 30:37:30.7 & 13.55 & 0.27 & 1.67 & 1.1 & \\ 
101897 & 01:33:57.73 & 30:17:14.2 & 17.11 & 0.01 & 1.11 & 3.35 & M(cool LBV cand)\\ 
101914 & 01:33:57.74 & 30:37:31.0 & 13.56 & 0.26 & 1.74 & 1.1 & \\ 
102105 & 01:33:57.85 & 30:33:38.4 & 17.71 & 0.07 & 1.32 & 1.27 & m(B2:I\_B467)\\ 
102367 & 01:33:58.00 & 30:41:22.3 & 15.90 & 0.23 & 1.49 & 1.12 & \\ 
102526 & 01:33:58.08 & 30:33:28.6 & 17.76 & 0.12 & 1.5 & 1.55 & \\ 
103164 & 01:33:58.43 & 30:33:01.6 & 18.06 & 0.28 & 1.48 & 1.27 & \\ 
103667 & 01:33:58.69 & 30:35:26.5 & 16.47 & $-$0.07 & 2.6 & 1.52 & m(B1Ia+WNE\_M33WR115=OB2$-$4),M\\ 
104242 & 01:33:59.01 & 30:33:53.9 & 17.35 & $-$0.02 & 1.46 & 1.4 & M(B8I in HII),UIT278\\ 
104285 & 01:33:59.03 & 30:33:56.7 & 18.42 & 0.05 & 2.41 & 1.1 & UIT269\\ 
104432 & 01:33:59.11 & 30:34:37.2 & 17.88 & 0.04 & 5.84 & 1.57 & star in HII\\ 
104958 & 01:33:59.40 & 30:23:11.0 & 16.95 & 0.03 & 1.28 & 2.12 & m(A0Ia\_IFM$-$B1330),M(HII)\\ 
105428 & 01:33:59.67 & 30:33:33.1 & 18.07 & 0.12 & 3.07 & 1.55 & HBS236226,M\\ 
105786 & 01:33:59.88 & 30:33:54.9 & 16.06 & 0.01 & 6.53 & 4.55 & \\ 
105799 & 01:33:59.89 & 30:34:27.2 & 17.91 & $-$0.11 & 5.14 & 1.6 & m(B\_ob4$-$39),M\\ 
106177 & 01:34:00.10 & 30:46:15.0 & 17.52 & 0.17 & 1.1 & 3.25 & \\ 
106653 & 01:34:00.41 & 30:37:18.6 & 18.36 & 0.14 & 1.35 & 1.1 & SFZ328,HBS236446\\ 
107707 & 01:34:01.05 & 30:36:34.7 & 18.12 & 0.12 & 2.63 & 1.2 & \\ 
107775 & 01:34:01.08 & 30:36:19.6 & 17.35 & 0.05 & 4.73 & 2.62 & UIT280,HBS236592,M\\ 
108708 & 01:34:01.68 & 30:37:20.0 & 18.40 & $-$0.08 & 5.55 & 1.25 & SFZ337,M(HII)\\ 
109058 & 01:34:01.92 & 30:38:19.3 & 17.84 & $-$0.10 & 3.05 & 1.25 & M\\ 
109156 & 01:34:01.99 & 30:38:53.8 & 17.43 & 0.04 & 2.88 & 2.27 & M\\ 
109457 & 01:34:02.21 & 30:38:50.3 & 18.12 & $-$0.06 & 5.6 & 1.38 & M\\ 
109749 & 01:34:02.43 & 30:31:03.5 & 18.43 & $-$0.02 & 2.26 & 1.3 & SFZ349,HBS236868,M\\ 
115127 & 01:34:06.63 & 30:41:47.8 & 15.95 & 0.07 & 7.25 & 2.33 & m(LBVcand),UIT301,M\\ 
115225 & 01:34:06.72 & 30:41:54.5 & 18.47 & $-$0.18 & 7.74 & 1.4 & M,m(EarlyO),UIT302,M\\ 
115375 & 01:34:06.83 & 30:47:22.4 & 17.46 & 0.32 & 14.86 & 1.57 & \\ 
116383 & 01:34:07.70 & 30:45:22.9 & 18.46 & 0.28 & 1.73 & 1.25 & SFZ369\\ 
116517 & 01:34:07.82 & 30:47:31.9 & 18.28 & $-$0.02 & 2.95 & 1.15 & HBS130240,M\\ 
117163 & 01:34:08.34 & 30:34:59.2 & 18.60 & 0.04 & 1.25 & 1.38 & \\ 
119710 & 01:34:10.61 & 30:26:00.5 & 18.14 & 0.04 & 1.15 & 1.4 & SFZ393,HBS220620,M\\ 
120082 & 01:34:10.93 & 30:34:37.6 & 16.22 & 0.12 & 3.31 & 1.1 & SFZ394,M(hot LBV cand),var\,83\\ 
120141 & 01:34:10.99 & 30:46:33.0 & 18.06 & 0.08 & 1.28 & 1.15 & M\\ 
120145 & 01:34:10.99 & 30:24:55.0 & 13.09 & 0.12 & 2.21 & 1.73 & \\ 
120530 & 01:34:11.33 & 30:36:28.0 & 18.63 & 0.10 & 4.12 & 1.55 & SFZ543,HBS220732,M\\ 
120786 & 01:34:11.56 & 30:36:25.4 & 17.92 & $-$0.00 & 4.07 & 2.17 & HBS220767,M\\ 
120805 & 01:34:11.58 & 30:36:17.4 & 18.32 & 0.03 & 3.83 & 1.32 & HBS220771,M\\ 
123649 & 01:34:14.21 & 30:53:55.2 & 18.32 & 0.10 & 2.77 & 2.62 & SFZ405\\ 
123651 & 01:34:14.21 & 30:33:43.3 & 18.49 & $-$0.03 & 8.86 & 1.2 & HBS221094,M(HII)\\ 
124598 & 01:34:15.03 & 30:34:57.8 & 18.16 & 0.22 & 3.07 & 1.2 & UIT330,SFZ550,HBS221214,M \\ 
124844 & 01:34:15.22 & 30:36:59.0 & 18.20 & 0.18 & 1.22 & 1.05 & M\\ 
125294 & 01:34:15.56 & 30:37:12.6 & 15.82 & 0.03 & 10.02 & 3.52 & \\ 
125342 & 01:34:15.61 & 30:41:11.0 & 18.59 & $-$0.23 & 1.36 & 1.27 & SFZ415,HBS131495,M\\ 
125850 & 01:34:16.05 & 30:33:44.2 & 16.87 & 0.02 & 4.01 & 1.18 & UIT341,SFZ423,M\\ 
125867 & 01:34:16.07 & 30:36:42.1 & 18.09 & $-$0.04 & 6.59 & 1.73 & UIT339,SFZ422,HBS221349,M(P Cyg LBV cand)\\ 
125919 & 01:34:16.10 & 30:33:44.9 & 16.56 & 0.06 & 2.35 & 1.2 & UIT341,SFZ423,m(LBVcand=B526),M\\ 
126188 & 01:34:16.35 & 30:37:12.3 & 18.17 & $-$0.18 & 9.07 & 1.35 & HBS221391,M,m(WN7\_M33WR130),UIT343,HII\\ 
127134 & 01:34:17.20 & 30:33:39.2 & 17.61 & $-$0.07 & 5.53 & 1.88 & HBS221507,M\\ 
127502 & 01:34:17.56 & 30:33:39.3 & 17.72 & $-$0.02 & 4.63 & 1.55 & HBS221547,M\\ 
128224 & 01:34:18.36 & 30:38:36.9 & 18.12 & $-$0.06 & 4.31 & 1.8 & SFZ439,HBS221640,M\\ 
130073 & 01:34:20.68 & 30:48:59.9 & 17.97 & 0.09 & 1.76 & 1.62 & \\ 
130074 & 01:34:20.68 & 30:39:42.7 & 18.14 & 0.01 & 1.83 & 0.9 & SFZ450,HBS221885,M\\ 
130270 & 01:34:20.95 & 30:30:39.9 & 16.59 & 0.17 & 1.1 & 2.98 & \\ 
132716 & 01:34:24.78 & 30:33:06.6 & 17.02 & 0.14 & 1.05 & 1.4 & M(cool LBV cand)\\ 
134181 & 01:34:27.11 & 30:45:59.8 & 17.93 & 0.03 & 2.96 & 2.23 & SFZ465,M\\ 
135855 & 01:34:29.71 & 30:53:12.0 & 18.16 & 0.29 & 1.37 & 1.18 & \\ 
136261 & 01:34:30.29 & 30:40:39.8 & 17.82 & $-$0.03 & 1.09 & 1.25 & M(star in HII)\\ 
137219 & 01:34:31.97 & 30:46:49.8 & 18.56 & $-$0.14 & 6.46 & 1.15 & m(EarlyO),UIT361,M\\ 
139873 & 01:34:37.28 & 30:38:17.8 & 17.61 & 0.33 & 1.17 & 1.38 & SFZ495\\ 
143582 & 01:34:49.96 & 30:29:21.3 & 13.86 & 0.32 & 1.24 & 0.85 & \\ 
144082 & 01:34:52.76 & 30:28:12.2 & 13.44 & 0.14 & 1.81 & 1.02 & \\ 
145023 & 01:34:59.39 & 30:42:01.2 & 18.43 & $-$0.13 & 1.21 & 1.0 & SFZ511HBS120967,m(O8Iaf\_ob88$-$7),M \\ 
145038 & 01:34:59.47 & 30:37:01.9 & 18.58 & 0.26 & 10.43 & 1.12 & SFZ512,HBS210675,M(hot LBV cand)\\ 
146074 & 01:35:09.73 & 30:41:57.3 & 18.15 & $-$0.22 & 5.09 & 1.1 & HBS110031,M(Ofpe/WN9 Romano's Star) \\ 

\end{longtable}

\begin{longtable}{c|c|c|c|c|c|c|l}
\caption{LBV candidates with $V >$16\fm0 and
\BVlim{}$< B-V <$1\fm{}2. Designations are similar to those used in
Table~\ref{FullList}}
\label{FullListRed}\\

\hline%\hline
N & $\alpha$ &  $\delta$ & V & $B-V$ & s &R($''$) & comment\\
\hline
1 & 2 & 3 & 4 & 5 & 6 & 7 & 8 \\
\hline
\endfirsthead
\caption{(Contd.)}\\

\hline
1 & 2 & 3 & 4 & 5 & 6 & 7 & 8 \\
\hline
\endhead

\hline
\endfoot

\hline%\hline
\endlastfoot

002627 & 01:32:31.94 & 30:35:16.7 & 17.47 & 0.56 & 2.12 & 1.1 & \\
006862 & 01:32:42.26 & 30:21:14.1 & 17.33 & 0.52 & 14.07 & 1.6 & M(hot LBV)\\
009835 & 01:32:48.06 & 30:24:50.3 & 18.07 & 0.76 & 2.33 & 1.18 & \\
021266 & 01:33:03.40 & 30:30:51.2 & 17.49 & 0.59 & 3.33 & 1.18 & \\
031584 & 01:33:14.81 & 30:45:59.3 & 18.47 & 0.5 & 2.72 & 1.35 & \\
045901 & 01:33:27.40 & 30:30:29.5 & 17.51 & 0.5 & 4.76 & 1.3 & \\
052581 & 01:33:31.80 & 30:22:59.1 & 16.27 & 0.66 & 8.95 & 1.73 & \\
057412 & 01:33:34.61 & 30:40:56.2 & 18.08 & 0.57 & 3.4 & 1.38 & \\
058746 & 01:33:35.33 & 30:41:46.1 & 17.39 & 0.67 & 4.9 & 1.88 & \\
061849 & 01:33:37.04 & 30:36:37.6 & 16.38 & 0.46 & 3.12 & 1.05 & \\
073136 & 01:33:42.52 & 30:32:58.6 & 17.67 & 0.4 & 6.58 & 1.12 & \\
075866 & 01:33:43.71 & 30:39:05.1 & 17.65 & 0.39 & 7.15 & 1.35 & \\
077435 & 01:33:44.40 & 30:44:28.1 & 17.75 & 0.56 & 3.29 & 1.35 & \\
077731 & 01:33:44.52 & 30:44:32.3 & 18.47 & 0.41 & 12.29 & 1.07 & M(HII/OB+neb)\\
078101 & 01:33:44.67 & 30:36:11.4 & 17.63 & 1.18 & 2.14 & 1.15 & \\
079224 & 01:33:45.15 & 30:36:20.1 & 16.59 & 0.4 & 2.01 & 1.12 & \\
086876 & 01:33:48.89 & 30:21:48.6 & 17.73 & 0.74 & 2.56 & 1.32 & \\
088927 & 01:33:49.94 & 30:29:28.8 & 18.23 & 0.78 & 2.38 & 1.38 & M(HII)\\
104139 & 01:33:58.96 & 30:41:39.5 & 16.93 & 0.51 & 3.61 & 1.07 & \\
115963 & 01:34:07.32 & 30:47:32.4 & 18.58 & 0.35 & 7.57 & 1.18 & HBS130137,m(HII)\\
124485 & 01:34:14.93 & 30:34:36.4 & 18.49 & 0.77 & 2.02 & 1.93 & HBS221202\\
124864 & 01:34:15.23 & 30:37:04.9 & 17.68 & 0.84 & 4.31 & 1.0 & \\
125093 & 01:34:15.42 & 30:28:16.4 & 17.47 & 0.73 & 4.22 & 1.12 & \\
141751 & 01:34:42.14 & 30:32:16.0 & 17.56 & 0.77 & 3.24 & 1.4 & M(hot LBV cand?)\\
146528 & 01:35:23.21 & 31:06:38.5 & 17.63 & 0.75 & 10.41 & 1.35 & \\

\end{longtable}

\end{document}